\documentclass[twocolumn,aps,prl,showpacs,floatfix,superscriptaddress]{revtex4}
\usepackage{amsmath}
\usepackage{amsfonts}
\usepackage{amssymb}
\usepackage{bm}
\usepackage{color}
\usepackage{graphicx}

\newcommand{\la}{\langle}
\newcommand{\ra}{\rangle}

\newcommand{\fdt}{\frac{d}{dt}}
\newcommand{\AT}{{\bf A}^T}
\newcommand{\ST}{{\bf S}^T}

\newcommand{\bA}{{\bf A}}
\newcommand{\bS}{{\bf S}}

\newcommand{\bB}{{\bf B}}

\newcommand{\Tr}{\mbox{Tr}}
\newcommand{\ee}{\end{equation}}
\newcommand{\be}{\begin{equation}}

\begin{document}
\title{Multi-scale model of gradient evolution in turbulent flows}

\author{Luca Biferale} \affiliation{Dip. Fisica and INFN,
  Universit\`a di ``Tor Vergata'' Via della Ricerca Scientifica
  1, 00133 Roma, Italy.}

\author{Laurent Chevillard} \affiliation{Department of Mechanical
  Engineering and Center for Environmental and Applied Fluid
  Mechanics, The Johns Hopkins University, 3400 N. Charles Street,
  Baltimore, MD 21218, USA}

\author{Charles Meneveau} \affiliation{Department of Mechanical Engineering
  and Center for Environmental and Applied Fluid Mechanics, The Johns
  Hopkins University, 3400 N. Charles Street, Baltimore, MD 21218,
  USA}

\author{Federico Toschi} \affiliation{Istituto per le Applicazioni del
  Calcolo CNR, Viale del Policlinico 137, 00161 Roma, Italy\\ and INFN,
  Sezione di Ferrara, Via G. Saragat 1, I-44100 Ferrara, Italy.}

\date{\today}
\begin{abstract}
  A multi-scale model for the evolution of the velocity gradient
  tensor in fully developed turbulence is proposed. The model is based
  on a coupling between a ``Restricted Euler'' dynamics [{\it
    P. Vieillefosse, Physica A, {\bf 14}, 150 (1984)}] which
  describes gradient self-stretching, and a deterministic
  cascade model which allows for energy exchange between different
  scales. We show that inclusion of the cascade process is sufficient
  to regularize the well-known finite time singularity of the
  Restricted Euler dynamics. At the same time, the  model retains
  topological and geometrical features of real turbulent flows: these
  include the alignment between vorticity and the intermediate
  eigenvector of the strain-rate tensor and the typical teardrop
  shape of the joint probability density between the two invariants,
  $R-Q$, of the gradient tensor. The model also possesses skewed,
  non-Gaussian longitudinal gradient fluctuations and the correct
  scaling of energy dissipation as a function of Reynolds
  number. Derivative flatness coefficients are in good agreement
  with experimental data.
\end{abstract}

\maketitle

The spatio-temporal fluctuations of small-scales in three-dimensional
turbulent flows are among the most complex phenomena known to
classical physics, being both highly non-Gaussian and strongly
long-range correlated. For example, velocity gradients as well as
velocity increments between two points typically show strong
fluctuations much larger than their standard deviation
\cite{sreene,frisch}. The same is true for fluid accelerations or
velocity increments at two different times
\cite{boden_acc,bife_acc,frisch}, with long-range correlations up to
the time scales of the largest eddies in the flow
\cite{long_range_acc}.  A possible mechanism for the large
fluctuations is the nonlinear self-stretching \cite{vieille,cantwell}
that occurs during the Lagrangian evolution of the velocity gradients,
$A_{ij} =\partial_i u_j$. This local self-stretching must be coupled
with the energy exchange among larger/smaller eddies and with velocity
fluctuations at different spatial locations via the pressure term. The
non-linear coupling among different scales also relates to the concept
of energy cascade, often invoked to explain the growth of
non-Gaussianity going from large to small scales.\\
Significant advances in experimental techniques now allow to
measure all components of ${\bf A}$ in different spatial locations
\cite{tsinober,tao,zeff} in high Reynolds number flows. These
experimental measurements, together with data generated using
direct numerical simulations, has uncovered the existence of many
interesting, and possibly universal, geometric features of ${\bf
A}$. Namely, (i) the preferred alignment of the vorticity vector
with the eigenvector of the intermediate eigenvalue of the
strain-rate tensor, ${\bf S} = ({\bf A} + {\bf A}^T)/2$; (ii) the
axisymmetric character of local deformation (two positive and one
negative eigenvalue of ${\bf S}$); (iii) the typical teardrop
shape of the joint probability distribution, $P(R,Q)$ where $R=
-\Tr({\bf A}^3)/3$ and $Q = -\Tr({\bf
  A}^2/2)$ are two invariants of ${\bf A}$
\cite{cantwell,tsinober,lund,vanderbros,pumir}.

A systematic analysis of the dynamics of velocity gradients was
made by Vieillefosse \cite{vieille}.  He started from the exact
equations governing the Lagrangian evolution of ${\bf A}$ in the
incompressible Navier-Stokes (NS) equations: $ \frac{d}{dt} A_{ij}
= -A_{ik}A_{kj} -\partial_i\partial_j p +\nu \partial^2 A_{ij},
\label{eq:ns} $ where $p$ is the pressure divided by density,
$\nu$ is the kinematic viscosity and with $d/dt$ we mean the
Lagrangian derivative.  Then, he retained only the isotropic part
of the pressure Hessian, $\partial_i
\partial_j p \sim \delta_{ij} \partial^2 p$ and used the
imcompressibility condition to express the pressure Laplacian in terms
of $\bA$. Finally he neglected the viscous contribution arriving to
the closed ``Restricted Euler'' (RE) equations:
\begin{equation}
\label{eq:vielle}
\fdt \bA = -\bA^2 +\Tr(\bA^2){\bf I}/3.
\end{equation}
It is remarkable that such a simple system is already sufficient
to explain many of the geometrical trends found in the real
gradient evolution, as shown by \cite{vieille,cantwell}. On the
other hand, the self-stretching mechanism is not constrained by
any energy exchange/loss mechanism, leading to a finite time
singularity for any initial condition. This blow up prevents the
use of the RE dynamics to make any systematic assessment of the
gradient's stationary statistics. Prior models that seek to
regularize the RE dynamics include a stochastic model with
log-normal statistics of the dissipation \cite{girimaji_pope}, a
linear and non-linear damping model for the viscous term
\cite{martin}, and a model where the material deformation history
described by a tetrad of points plus some stochastic terms are
used to mimic the anisotropic pressure fluctuations \cite{pumir}.
Recently, a model of the anisotropic pressure Hessian and of the
viscous term has been proposed by introducing a finite-time memory
effect in the closure of the material deformation tensor
\cite{cm}. Results show that this model reproduces stationarity,
non-Gaussianity of transverse and longitudinal gradients and the
correct geometrical properties of ${\bf A}$. Remarkably, the
system also predicts experimentally observed relative scaling
exponents characterizing how non-Gaussian statistics evolves with
Reynolds number, at least from small to moderate values of it.
However, attempts to simulate velocity gradient dynamics at
arbitrarily large Reynolds numbers with this model have been, so
far, unsuccessful. It was concluded that the difficulties were
associated to the assumptions that the velocity gradient tensor
evolves mainly uncoupled from larger/smaller eddies and
neighboring locations.
\begin{figure}[!t]
\includegraphics[width=\hsize]{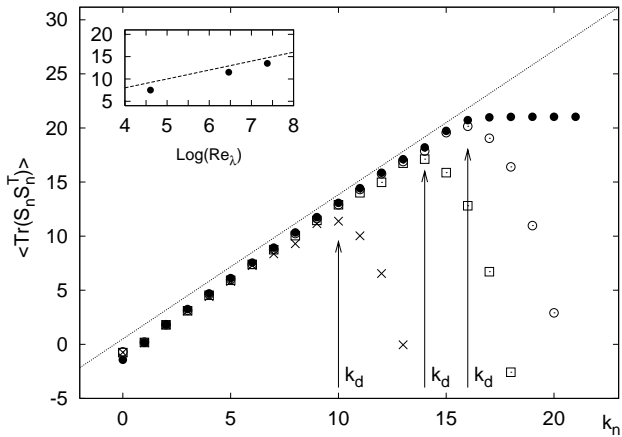}
\caption{Log-log plot of the mean energy dissipation (divided by
  $2\nu$): $\la \Tr ({\bf S}_n {\bf S}_n^T) \ra $ versus $k_n$, for
  $Re_{\lambda}=1,500$ ($\circ$); $Re_{\lambda}=600$ ($\Box$); $Re_{\lambda}=130$     ($\times$) . The dissipative scale $k_d$  is defined as the
  wavenumber where the energy dissipation peaks (arrows in the
  plot). The straight line correspond to the Kolmogorov spectrum $
  \sim k_n^{4/3}$. The data which saturates in the viscous range,
  $\bullet$, are for the coarse-grained (defined later in the text)
   variables: $\la \Tr ({\bf S}_n^{CG}  {{\bf S}_n^{CG}}^T) \ra $.
   In the inset we show the Reynolds
  dependency of the energy dissipation evaluated at $k_d$.
The straight line corresponds  to the expected
slope $\propto Re_{\lambda}^{2}$. }
\label{fig:1new}
\end{figure}
In this paper we propose a minimal model in which deterministic
couplings among scales of motion are included. The model couples
the self-stretching local dynamics {\it \`a la} Vieillefosse with
a cascade mechanism, i.e. by coupling the velocity evolution on
the gradient scale given by (\ref{eq:vielle}) with velocity
fluctuations at larger scales. The most remarkable result is that
such a coupling will be shown to be sufficient to regularize the
Vieillefosse finite-time singularity, without destroying the main
positive features of (\ref{eq:vielle}) including the presence of
skewed longitudinal gradients and long tails in the probability
density functions of gradients; the vorticity alignment and the
typical tear-drop shape of the joint probability, $P(R,Q)$. Also
the intermittency level as measured by the flatness coefficient is
in good quantitative agreement with experiments
\cite{antonia81,sreene}.  The model thus provides meaningful
results even for very large Reynolds number (here
we present results up to $Re_{\lambda}=1,500$).\\
To introduce a cascade dynamics we start from a decomposition of
the gradient tensor into band-passed contributions $\bA = \sum_n
\bA_n$, each $\bA_n$ describes the velocity gradient at a typical
wavenumber $k_n$ \cite{Greg}.  The set of possible wavenumbers is
chosen equispaced on a logarithmic scale, $k_n =2^n k_0$, such as
to optimally capture power-law behaviors in the inertial range.
The band-passed version of the NS equations is of the form
\begin{equation}
\label{eq:nsshell}
\fdt \bA_n = -\sum_{p,q} \left(\bA_p \bA_q\right)_n +
\bB_n  - \frac{\Tr(\%)}{3} {\bf I} + \nu \partial^2  \bA_n ,
\end{equation}
where the trace term is added to keep the entire right-hand-side
trace-free, and the term $\bB_n$ represents pressure effects,
interscale interactions and additional spatial transport
introduced by the band-pass filtering acting on the advective term
on the left side of the original equation. Neglecting the viscous
term and $\bB_n$, and keeping only the fully diagonal term
($p=q=n$) in the double sum leads to the most severe
approximation. Namely that the gradients, $\bA_n$, band-passed on
different shells, follow the RE dynamics separately
shell-by-shell: $d\bA_n/dt = -\bA_n^2 +1/3 {\Tr(\bA_n^2)}{\bf I}$.
Of course this set of uncoupled equations suffers from the same
drawbacks of the original Vieillefosse model.  The coupling
between different scales must be introduced by keeping some terms
besides the purely diagonal one in the first term in the rhs of
Eq. \ref{eq:nsshell}: i.e. we include a new term $F$ and we choose
its form by imposing an energy preserving structure for the
coupling terms, following the assumption that the pressure-Hessian
does not perform work on the band-passed gradient \cite{pumir}.
Moreover, it is natural to suppose that the most relevant
dynamical interactions happen within nearby scales, (locality
assumption) and therefore we limit the range of interactions to
the next-nearest neighbors (up to $n \pm 2$, for each $k_n$). 
\begin{figure}[!t]
\begin{center}
  \includegraphics[width=\hsize]{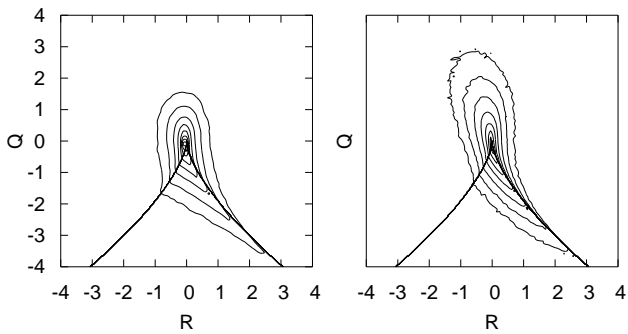}
  \caption{Isolines of $P(R,Q)$ for DNS and model data. Right panel:
    DNS data at $Re_{\lambda} \sim 180$.  Left panel: model data at
    $Re_{\lambda}=140$, evaluated at the dissipative scale $k_d(Re)$.
    The continuous line in all diagrams correspond to the Vieillefosse
    zero-discriminant curve $Q^3+27/4 R^2=0$.}
\label{fig:3new}
\end{center}
\end{figure}
One
possible choice for the coupling term on the $n$th shell is:
\begin{eqnarray}
  &&{\bf F}_n[ A,  A ] =  \bA_{n+2}\AT_{n+1} + b2^2  \AT_{n-1}\bA_{n+1} + \nonumber \\
  &&(1-b) 2^4 \bA_{n-2} \bA_{n-1} -\Tr[\%],
  \label{eq:sabra}
\end{eqnarray}
where $b$ is a free parameter (always fixed to $b=0.5$ hereafter). Let
us notice that the structure of ${\bf F}_n[A,A]$ is chosen such that
the total energy  $E = \sum_n k_n^{-2} \Tr(\bA_n\AT_n)$ is
preserved by the effects of the three nonlinear terms. The model form
is proposed as a model for the neglect of all the truncated terms in
the double sum and the $\bB_n$ transport term.  The functional form
proposed for ${\bf F}_n[A,A]$ is motivated by the typical structure
of the nonlinear terms used in Shell models of turbulence
\cite{shell}, with the important difference that via (\ref{eq:sabra})
we can access now the whole geometrical properties of the band-passed
variables, owing to the tensorial structure of the $\bA_n$ variables.
To introduce a characteristic wavenumber from which the gradient variables
receive their maximal contribution, we need a viscous damping.  The simplest dimensionally consistent way to do this is by modeling the
viscous Laplacian term in NS equations using a linear damping term on
each band proportional to $\nu k_n^2$, where $\nu$ is the
viscosity. Combining all these elements one obtains the following dynamical system:
\begin{equation}
  \label{eq:vielleshellmatrix}
  {{d \bA_n}\over {dt}} = \alpha \left[-\bA_n^2+\frac{1}{3}\Tr(\bA_n^2){\bf I})\right] +(1-\alpha) \left({\bf F}_n   -\nu k_n^2 \bA_n\right)
\end{equation}
where $\alpha$ is a parameter which weights the relative importance of
the RE dynamics with respect to the energy exchange and energy
dissipation terms. As a first step, in this work we restrict attention
to the case $\alpha =0.5$. The system consist of $9 N$ coupled
nonlinear equations.  In the previous equations the infrared and
ultraviolet truncation is imposed by keeping in the nonlinear term
$A_n=0$ for $n=\{-1,-2, N+1, N+2\}$.

\begin{figure}[!t]
\includegraphics[width=\hsize]{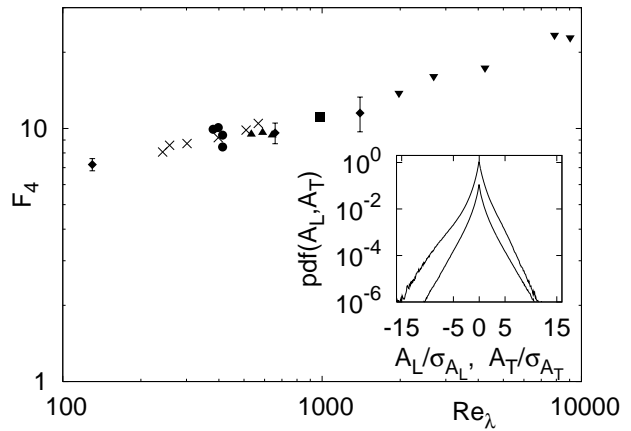}
\caption{ Comparison between a collection of experimental data for the
  longitudinal gradient flatness \cite{antonia81}, superposed with the
  results from our model (diamonds with error bars). Where we have
  used the variance of the longitudinal coarse grained gradient,
  $\sigma_A =\la (A_{11}^{CG})^2 \ra^{1/2}$, to define the Reynolds
  number: $Re_{\lambda} = 2E/(3 \nu \sigma_A)$ \cite{frisch}.  Inset:
  Data for $Re_{\lambda} = 1500$.  Normalized Pdf of longitudinal
  (top) and transverse (bottom) band-passed gradients measured at
  $k_d$.  Notice the skewed profile for the longitudinal
  gradients. Longitudinal pdf has been shifted along the y-axis for
  sake of clarity.}
\label{fig:2new}
\end{figure}

The main question we want to answer now is if a stationary statistical
state can be achieved once a forcing is applied at large scales. In
other words, we want to understand if the simple energy-exchange term,
${\bf F}_n[A,A]$, is sufficient to regularize the dynamics of the RE
structure at all scales.  Hopefully, this will allow us also to study
the important questions related to the dependency of geometrical
structure of turbulence on the Reynolds numbers and its correlation
with inertial range quantities.  The simplest way to achieve a
stationary state is to add a white-in-time Gaussian variable, ${\bf
  G}$, at the largest band, $n=0$, where in order to satisfy the
requirements of homogeneity and isotropy, we must chose the covariance
of the forcing to be $\la
G_{ij}G_{ml} \ra =  2 \delta_{im}\delta_{jl} -1/2 \delta_{ij}\delta_{ml}-1/2 \delta_{il}\delta_{jm}$ \cite{cm}.\\
The gradient statistics will be studied by looking at the shell $k_d$
(where $n=d$), defined as where the spectrum of the $\bA_n$ has its
peak, i.e. where a balance between the quadratic terms and the viscous
terms of (\ref{eq:vielleshellmatrix}) is obtained.  The system of
equations is integrated numerically using a 4-th order Runge-Kutta
scheme.  Three Reynolds numbers are simulated by using viscosities
equal to $\nu = \{10^{-4}, 2.7 \times 10^{-6}, 4.4 \times10^{-7}\}$
and with a total number of shells $N=\{14, 18, 22\}$, corresponding to
$Re_{\lambda} = \{130, 640, 1,500\}$ respectively.  Results are shown
in Fig \ref{fig:1new}, which shows the time-averaged energy
dissipation (without multiplication by $\nu$), measured on the
band-pass gradient variables, $ \la \Tr (\bA_n \AT_n) \ra $ as
function of $k_n$: notice that it has the expected Kolmogorov scaling
$\la \Tr (\bA_n \AT_n) \ra \sim \la \Tr ({\bS}_n \ST_n) \ra \sim k_n
^{4/3}$.
\begin{figure}[!t]
\begin{center}
\includegraphics[width=\hsize]{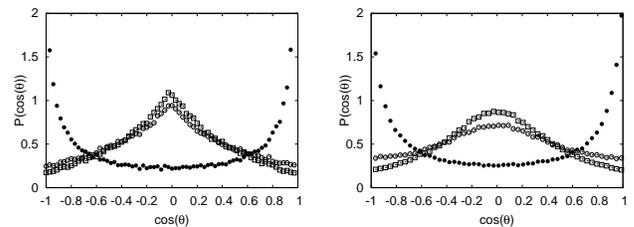}
\caption{PDF of the cosine between the vorticity and the eigenvectors
  associated with the Intermediate ($+$), maximum ($\star$) and
  smallest ($\times$) eigenvalues of the strain-rate matrix.  Notice
  the good agreement between the DNS (right) data and the model data
  (left).}
\label{fig:4new}
\end{center}
\end{figure}
On dimensional grounds we expect that the wavenumber where the
matching is achieved scales with Reynolds number as $k_d \sim
Re_{\lambda}^{2}$. This is indeed verified in the inset of Fig
(\ref{fig:1new}).  Geometrical properties of small-scale turbulent
fluctuations can be monitored by studying the joint statistics of the
two invariants, $Q,R$ at different Reynolds number and/or at different
scales. Here, by measuring the simultaneous distribution of $Q_n =
\Tr(\bA_n^2)$ and of $R_n = \Tr(\bA_n^3)$ for different Reynolds
numbers we can do both.  In Fig (\ref{fig:3new}) we show the isolines
of the joint probability distribution $P(R_n,Q_n)$ measured on the
dissipative scale $n=d$. We notice the presence of high probable
fluctuations along the right Vieillefosse tail, also seen in real
turbulence, which is shown in the same figure panel (b). The data are
from direct numerical simulation (DNS), done at a similar Reynolds
number.  As can be seen, there is very good agreement of the model
with the Navier-Stokes data, except for a small depletion of events in
the third quadrant ($R<0,Q>0$). In this quadrant, real data are
strongly affected by vortex stretching, an effect which is evidently
under-represented by the model evolution.  The $P(R,Q)$ distribution
has very little sensitivity to Reynolds number (at least in the range
investigated here). The $P(R,Q)$ distribution becomes more symmetric
for band-pass variables at larger scales (not shown), and this effect
is also observed in Navier-Stokes turbulence
\cite{vanderbros,pumir}. Finally, in Fig. (\ref{fig:4new}) we present
quantification of the alignment between the vorticity at the
dissipative scale, $(w_i)_d = \epsilon_{ijl} (A_{jl})_d$, and the
three eigenvectors of the strain rate tensor ${\bf S}_d$ also at that
scale.  It is well-known from DNS and experiments that vorticity tends
to preferentially align with the intermediate eigenvector
\cite{ashurst,tsinober,tao}: Figure (\ref{fig:4new}) shows that our
model is able to capture also this feature.  Finally, we document the
results in terms of intermittency. First, we show in the inset of Fig
(\ref{fig:2new}) the pdf of both longitudinal and transverse
gradients. Correctly, the model possesses longitudinal skewed
distribution while the transverse gradient pdf is fully symmetric. The
gradient fluctuations are highly non-Gaussian, as a results of the
growth of intermittency going from the large scale down to the
dissipative scale.  The growth of intermittency at decreasing length
scales is often characterized by measuring the flatness coefficient of
velocity gradients, $F_4 = {\langle A_d^4\rangle}/{\langle
  A_d^2\rangle^2}$ and plotting it as function of Reynolds number
$Re_{\lambda}$, where with $A_d$ we mean any of the longitudinal
components of $\bA$.  As a technical difficulty we point out that the
exponential fall-off of the gradient spectrum in the shell-matrix
model (\ref{eq:vielleshellmatrix}) prevents us from a direct
measurement of $F_4$ on the dissipative shell (the flatness tends to
grow exponentially in the dissipative range, making the estimate very
delicate).  For that reasons we decide to measure the flatness using a
coarse-grained variable. Specifically, the coarse-grained velocity
gradient at scale $k_m$ is defined as $A_m^{CG} = \sum_{n=0}^{n=m} A_n
$. For large $m$, we obtain a variable which includes fluctuations on
all scales, and which becomes independent on $m$ (saturates) beyond
the viscous range.  The two variables $A_n$ and $A_n^{CG}$ have
obviously the same scaling in the inertial range; in the dissipative
range the band-passed decays exponentially while the coarse-grained
saturates, see Fig (\ref{fig:1new}). This saturation allows us to have
robust measurements of intermittency in the dissipative range. In the
main body of Fig. (\ref{fig:2new}) we show the behaviour of Flatness
of the coarse-grained velocity gradient as a function of Reynolds
number, superposed with experimental data. Within the range of
Reynolds numbers considered, the agreement is very satisfactory.

In conclusion, we have introduced a ``shell'' version of the RE
dynamics which is free from the finite time singularity of the
original Vieillefosse formulation. The regularization is achieved
at all scales thanks to the introduction of an energy-exchange
mechanism with smaller and larger eddies. A viscous dissipative
term is introduced which acts preferentially at the viscous scale
where the gradients peak.  The model shows very realistic behavior
at changing Reynolds numbers, including (i) the skewed nature of
longitudinal gradients, (ii) the alignment of vorticity with the
intermediate eigenvector of the strain rate tensor, (iii) the
accumulation of events along the right tail of the Vieillefosse
line and (iv) the correct level of intermittency as measured by
the Flatness and its increasing trend at increasing
$Re_{\lambda}$.  A set of questions remains open.  From a
dynamical point of view, the model seems to under-predict the
probability of observing strong vortex stretching events - this
could point to open challenges associated with modeling
small-scale coherent vortices \cite{bife_acc}. From a statistical
point of view, it will be interesting to explore the effects of
varying the free parameters $\alpha$ and $b$ on intermittency
levels.  The free parameter $\alpha$ defines the relative
importance of the local Vieillefosse dynamics with respect to the
energy exchange mechanism.
 A promising development could be to couple the present multi-scale
 approach with the Lagrangian deformation method proposed in
 \cite{cm}, which proved to be successful in reproducing the main
 features of turbulent gradients at small and moderate Reynolds.\\
We thank R. Benzi for helpful discussions.
L.B. and F.T. acknowledge the hospitality of the Center for
Environmental and Applied Fluid Mechanics of Johns Hopkins University
where part of this work has been done. L.C. acknowledges the support
of the Keck Foundation, C.M. thanks the support of the US National
Science Foundation, F.T. thanks CNR for support under the Short Term Mobility
grant.


\begin{thebibliography}{99}
\bibitem{sreene} K.R. Sreenivasan and R. Antonia, {\it
    Annu. Rev. Fluid Mech.} {\bf 29} 435 (1997).
\bibitem{frisch} U. Frisch, {\it Turbulence: the legacy of
    A.N. Kolmogorov} (Cambridge University Press, Cambridge, 1995).
\bibitem{boden_acc} La Porta et al.  {\itshape Nature}~{\bfseries
    409}, 1017 (2001); {\itshape J. Fluid Mech.}~{\bfseries 469}, 121
  (2002).
\bibitem{bife_acc} L. Biferale et al. {\itshape
    Phys. Fluids.}~{\bfseries 17}, 021701 (2005); L. Chevillard et
    al., {\it Phys. Rev. Lett.} {\bf 91}, 214502 (2003).
\bibitem{long_range_acc} N. Mordant, E. Leveque and J-F.  Pinton,
  {\itshape New Journal of Physics} {\bf 6} 116 (2004).
\bibitem{vieille} P. Vieillefosse, {\it Physica A} {\bf 125} 150
  (1984).
\bibitem{cantwell} B.J. Cantwell, {\it Phys. Fluids A} {\bf 5}, 2008
  (1993); {\it Phys. Fluids A} {\bf 4}, 782 (1992)
\bibitem{tsinober}A. Tsinober, E. Kit and T. Dracos, {\bf 242} 169
  (1992); B. Luthi, A. Tsinober and W. Kinzelbach, {\it J. Fluid
    Mech.} {\bf 528} 87 (2005).
\bibitem{tao} B. Tao, J. Katz \& C. Meneveau {\it J. Fluid Mech.} {\bf
    467} 35 (2002).
\bibitem{zeff} B.W. Zeff et al., {\it Nature} {\bf 421} 146 (2003).
\bibitem{lund} T.S. Lund and M.M. Rogers, {\it Phys Fluids} {\bf 6}
  1838 (1994).
\bibitem{vanderbros} F. van der Bos et al., {\it Phys Fluids} {\bf 14}
  2457 (2002).
\bibitem{pumir} M. Chertkov, A. Pumir and B. Shraiman, {\it
    Phys. Fluids} {\bf 11} 2394 (1999); A. Naso and A. Pumir, {\it
    Phys. Rev. E} {\bf 72} 056318 (2005).
\bibitem{shell}L. Biferale, {\it Annu. Rev. Fluid Mech.} {\bf 35} 441
  (2003).
\bibitem{girimaji_pope} S.S. Girimaji abd S.B. Pope, {\it Phys Fluids
    A} {\bf 2} 242 (1990).
\bibitem{martin} J. Martin et al., {\it Phys. Fluids} {\bf 10} 2336
  (1998). E. Jeong and S. S. Girimaji,  {\it Theor. Comput. Fluid
  Dyn.}
{\bf 16}, 421 (2003).
\bibitem{cm} L. Chevillard and C. Meneveau, {\it Phys. Rev. Lett} {\bf
    97}, 174501 (2006).
\bibitem{Greg}
G. Eyink, {\it J. Fluid Mech.} {\bf 549},  159 (2006).


\bibitem{ashurst} W.T. Ashurst et al., {\it Phys. Fluids}~ {\bf
30},
  2343 (1987).
\bibitem{antonia81} R. A. Antonia, A. J. Chambers, and
  B. R. Satyaprakash {\it Boundary-Layer Meteorology} {\bf 21} (1981)
  159-171.
\end{thebibliography}
\end{document}